\documentclass[aps,amsmath,pr,twocolumn]{revtex4-1}
\usepackage{bm}
\usepackage{epsfig}
\usepackage{graphicx}
\usepackage{epstopdf}
\usepackage{amsmath}

\usepackage{color}
\usepackage{CJK}
\newcommand{\nix}[1]{}

\usepackage[unicode=true,colorlinks=true,citecolor=blue,urlcolor=blue]{hyperref}

\begin{document}
\begin{CJK*}{Bg5}{bsmi}
\title{Terahertz ratchet effects in graphene with a lateral superlattice}

\author{
P.\,Olbrich$^1$, 
J.\,Kamann$^1$, M.\,K\"{o}nig$^1$, J.\,Munzert$^1$, L.\,Tutsch$^1$, 
J.\,Eroms$^1$, D.\,Weiss$^1$, Ming-Hao\,Liu\,({¼B©ú»¨})$^1$, 
L.E.\,Golub$^2$, E.L.\,Ivchenko$^2$,  
V.V.\,Popov$^{3,4,5}$, D.V.\,Fateev$^{3}$, K.V.\,Mashinsky$^{4}$,
F.\,Fromm$^6$, Th.\,Seyller$^6$, 
and S.D.\,Ganichev$^{1}$
}
\email{e-mail: sergey.ganichev@physik.uni-regensburg.de}

\affiliation{$^1$ Physics Department, University of Regensburg,
93040 Regensburg, Germany}
\affiliation{$^2$ Ioffe Institute, 194021 St.~Petersburg, Russia}
\affiliation{$^3$ Institute of Radio Engineering and Electronics (Saratov Branch),
410019 Saratov, Russia} 
\affiliation{$^4$ Saratov State University, 410012 Saratov, Russia}
\affiliation{$^5$ Saratov Scientific Center of the Russian Academy of Sciences, 410028 Saratov, Russia}
\affiliation{$^6$ Technical University of Chemnitz, 09111 Chemnitz, Germany}

\begin{abstract}
Experimental and theoretical  studies on ratchet 
effects in graphene  with a lateral superlattice excited
by alternating electric fields of terahertz frequency range are 
presented. A lateral superlatice deposited on top of monolayer graphene
is formed either by periodically repeated metal stripes having different widths and spacings or
by inter-digitated comb-like dual-grating-gate (DGG) structures.
We show that the ratchet photocurrent excited by terahertz radiation
and sensitive to the radiation polarization state can be efficiently
controlled by the back gate driving the system through the Dirac point as well as
by the lateral asymmetry varied by applying unequal voltages to the DGG subgratings.
The ratchet photocurrent includes the 
Seebeck thermoratchet effect 
as well as the effects of ``linear'' and ``circular'' ratchets, sensitive to the 
corresponding polarization of the driving electromagnetic force.
The experimental data are analyzed for the electronic and plasmonic ratchets
 taking into account the calculated potential profile and the near field acting on carriers in graphene.
We show that the photocurrent generation is based on a combined
action of a spatially periodic in-plane potential and the spatially 
modulated light due to the near field effects of the light diffraction.
\end{abstract}


\maketitle
\end{CJK*}

\section{Introduction}

Graphene has revealed fascinating phenomena in a 
number of experiments owing to specifics of the electron energy 
spectrum resembling that of a massless relativistic 
particle~\cite{Review0,Review,Review2,Review3,Review4}. 
Unique physical properties of graphene, such as
the gapless linear energy spectrum, pure two-dimensional (2D) transport, 
strong plasmonic response,
and comparatively high mobility at room temperature,
open the prospect of high-speed electronics and optoelectronics, 
in particular, fast and sensitive detection of light for a 
range of frequencies from ultraviolet to terahertz (THz).
Different mechanisms, by which the detection can be accomplished, include: 
(i)~photoconductivity due to bolometric and photogating 
effects~\cite{bolometer,bolometer2,koppens}, 
(ii)~photo-thermoelectric (Seebeck) effect~\cite{photothermo}, (iii)~separation of the photoinduced electron-hole pairs in a 
periodic structure with two different metals  serving as contacts 
to graphene~\cite{mueller,weiss} (double comb structures)
or  a $p$-$n$ junction~\cite{novoselov}, and (iv)~excitation of 
plasma waves in a gated graphene sheet \cite{plasmawave,plasmawave2}, for reviews 
see~\cite{Tredicucci14,Koppens14,kono14,knaprev2013,knaptredicuccirev2013,Boppel2012,Tonouchi2007}.
As we show below graphene based detectors may operate applying ratchet effects
excited by THz radiation in a 2D crystal superimposed by 
a lateral periodic metal structure.  
The ratchet effect is of  very general nature: The spatially periodic, 
noncentrosymmetric systems being driven out of thermal equilibrium are able to 
transport particles even in the absence of an average macroscopic 
force~\cite{reimann}. This effect, so far demonstrated for semiconductor 
quantum wells with a lateral noncentrosymmetric 
superlattice structures~\cite{Olbrich_PRL_09,Olbrich_PRB_11,Review_JETP_Lett,popovAPL,otsuji,otsuji2,otsuji3,faltermeier15},
promises  high responsivity, short response times, and even
new functionalites, such as all-electric detection of the radiation polarization state
including radiation helicity being so far realized applying photogalvanics~\cite{helicitydetector,ellipticitydetector}. Most recently 
electronic and plasmonic ratchet effects have been considered 
theoretically in Ref.~\cite{Theory_PRB_11,popov2013,Budkin2014,Koniakhin2014,Rozhansky2015,Popov2015} supporting
the expected benefit of graphene based detectors.

In the present work, we report on the experimental realization and systematic study of  
graphene ratchets  in both (i) epitaxially grown and (ii) exfoliated graphene with an asymmetric lateral periodic potential. 
The modulated potential has been obtained by  fabrication of either a sequence of metal stripes  on top of graphene or
 inter-digitated comb-like dual-grating-gate structures. 
We demonstrate that  THz laser radiation shining on the modulated devices results in the excitation of a direct electric current being sensitive to the radiation's polarization state. The application of different electrostatic potentials  to  the two different subgratings of the dual-grating-gate structure and variation of the back gate potential 
enables us to change in a controllable way the degree and the sign of the structure asymmetry as well as to analyze the photocurrent behaviour upon changing the carrier type and density. These data reveal that the photocurrent reflects the degree of asymmetry induced by different top gate potentials and even vanishes for a symmetric profile. 
 Moreover, it is strongly enhanced in the vicinity of the Dirac point. 
The  measurements together with a beam scan across the lateral structure 
prove that the observed photocurrent stems from the ratchet effect.
The ratchet current consists of a few linearly independent contributions including the
Seebeck thermoratchet effect as well as the ``linear'' and ``circular'' ratchets, sensitive to the corresponding polarization of the driving electromagnetic force.
The results are analyzed in terms of the theory of ratchet effects in graphene
structures with a lateral potential   including the electronic and  plasmonic mechanisms of a 
photocurrent in periodic structures.
We show that the ratchet photocurrent appears due to the noncentrosymmetry of the periodic graphene structure unit cell. 
The experimental data and the theoretical model are discussed by taking 
the calculated potential profile and near-field effects explicitly into account.

\section{Samples and Methods}
 
\begin{figure}[t]
        \includegraphics[width=\linewidth]{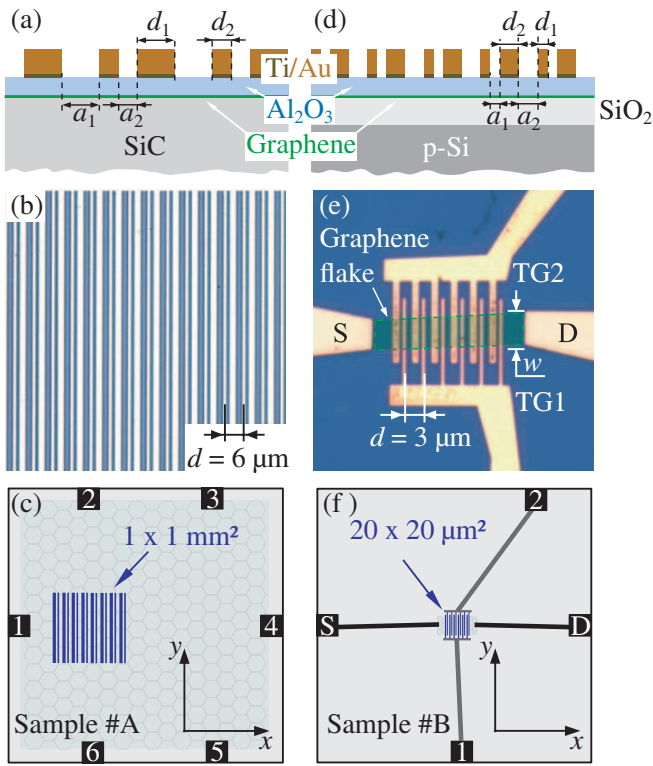}
\caption{Cross-section, photograph and sample geometry sketch of: 
(a)-(c) metal finger structure deposited on a large area epitaxial graphene (sample \#A) 
and (d)-(f) inter-digitated grating gates deposited on the exfoliated 
monolayer graphene flake (sample \#B). Here, $d_{1,2}$ and $a_{1,2}$ are the width of metal 
stripes and the spacing in-between, respectively. The superlattice period 
is $d= d_1+ d_2+a_1+a_2$.
}   
\label{fig_structure}
\end{figure}

We study the ratchet photocurrents in two different types of structures. 
The superlattices 
of the first type {are fabricated on} 
large area graphene grown by high-temperature Si sublimation of 
semi-insulating SiC substrates~\cite{Seyller1,Seyller2}.
This type of sample with the superlattice covering an area 
of about $1 \times 1$ mm$^2$ on a graphene layer with a total area of about 
$5 \times 5$ mm$^2$ allowed us, on one hand, to scan the laser beam 
across the superlattice and, on the other hand, 
to examine the photocurrent in directions along and perpendicular 
to the metal stripes. 
All samples are made from the same wafer of SiC.
To obtain defined graphene edges we removed an edge trim  of about 200 $\mu$m width, see Fig.~\ref{fig_structure}(c),
by reactive ion etching with an argon/oxygen 
plasma.
The carrier mobility $\mu = 1800$~cm$^2$/Vs and residual
hole density 5.3 $\times 10^{11}$cm$^{-2}$ 
in graphene  resulting in a carrier transport 
relaxation time $\tau$\,=\,16~fs were measured at $T = 200$~K. 
Before fabricating the superlattice structure, we carefully checked that 
the symmetry of the pristine graphene is unaffected by steps (terraces), which may be formed on the 
SiC surface.
For that we studied  the THz radiation induced photocurrent and ensured that it vanishes 
at normal incidence~\cite{footnote0,Karch_PRL_2010,Glazov2014}.
As a further step, we  deposited an insulating aluminium oxide layer on  top of the graphene sheet. 
For this purpose, we first  deposited a thin ($< 1$~nm) Al seed layer by evaporation 
in ultra high vacuum and oxidized it subsequently. Then we prepared 26~nm layer of aluminum oxide $Al_2O_3$ with atomic 
layer deposition using H$_2$O and trimethyl aluminum as precursors.
The lateral periodic electrostatic potential is created by micropatterned periodic grating-gate fingers  fabricated by
electron beam lithography and subsequent deposition of metal (5~nm Ti and 60 nm Au) on graphene covered  by Al$_2$O$_3$.
A sketch of the gate fingers and a corresponding optical micrograph are 
shown in Figs.~\ref{fig_structure}(a) and (b), respectively.
The grating-gate supercell consists of two metal stripes having different widths 
$d_1 = 2$~$\mu$m and $d_2 = 1$~$\mu$m separated by different spacings $a_1 = 2$~$\mu$m and $a_2 = 1$~$\mu$m.
This supercell is  repeated to generate an asymmetric periodic electrostatic potential~\cite{Olbrich_PRB_11,staab2015} (period $d = d_1 + a_1 +d_2 +a_2 = 6$~$\mu$m)
superimposed upon graphene, see Fig.~\ref{fig_structure}(b). 
The $1 \times 1$~mm$^2$ area grating-gate  structure is located on the left half of the sample 
so that a large graphene area remains unpatterned, see Fig.~\ref{fig_structure}(c).
For the THz beam of 1.5~mm diameter, this design allows us to study the 
photocurrent excited in either the superlattice structure or the unpatterned graphene reference area. 
Contact pads were placed in a way that the photo-induced currents 
can be measured parallel ($j_y$, contacts 2 and 6) and perpendicular ($j_x$, contacts 1 and 4) to the metal fingers.  
Two additional contacts (3 and 5) were used for detecting the photocurrent signals from the unpatterned area  as a reference.

\begin{figure}[t]
        \includegraphics[width=\linewidth]{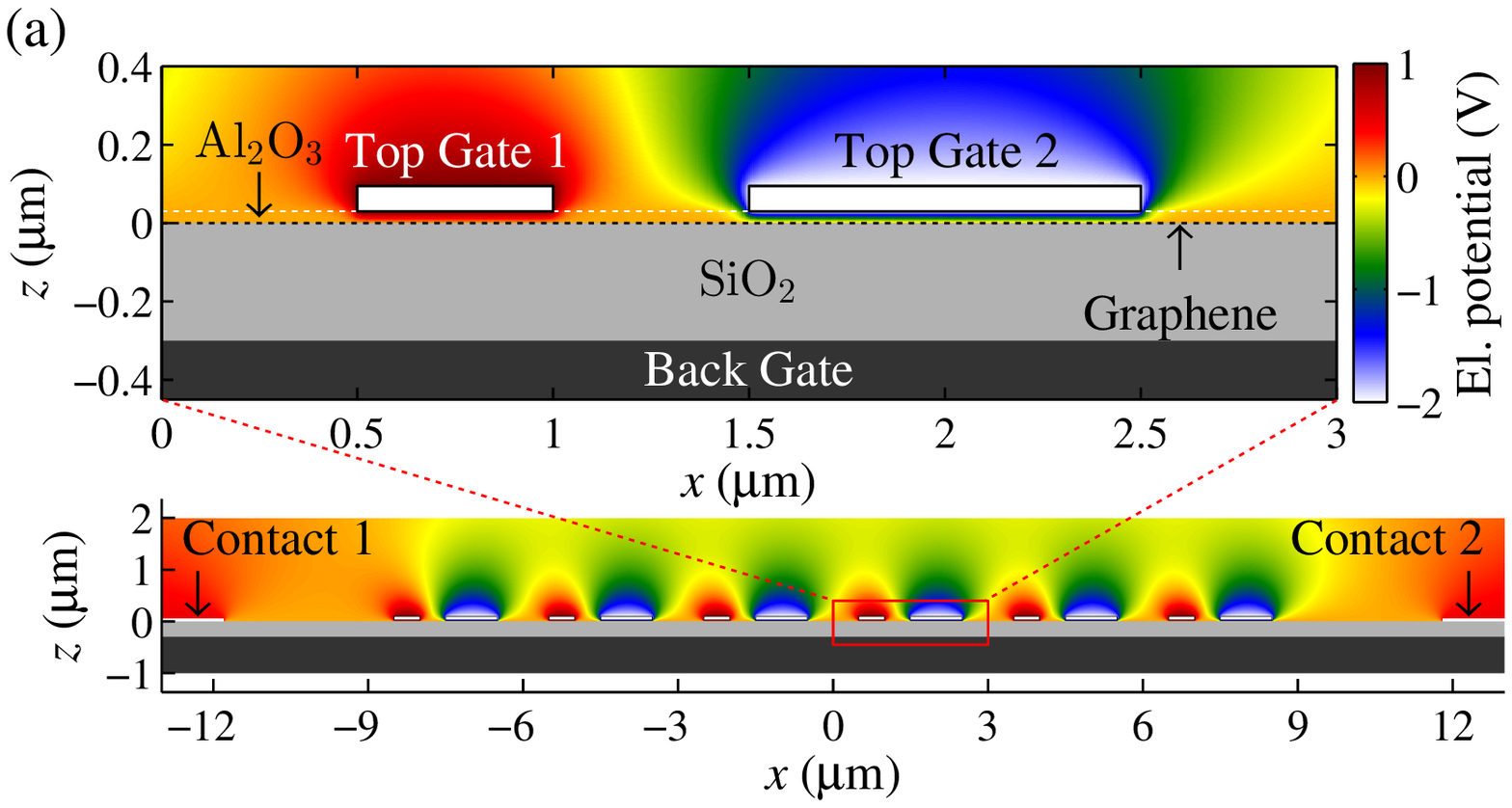}
        \includegraphics[width=\linewidth]{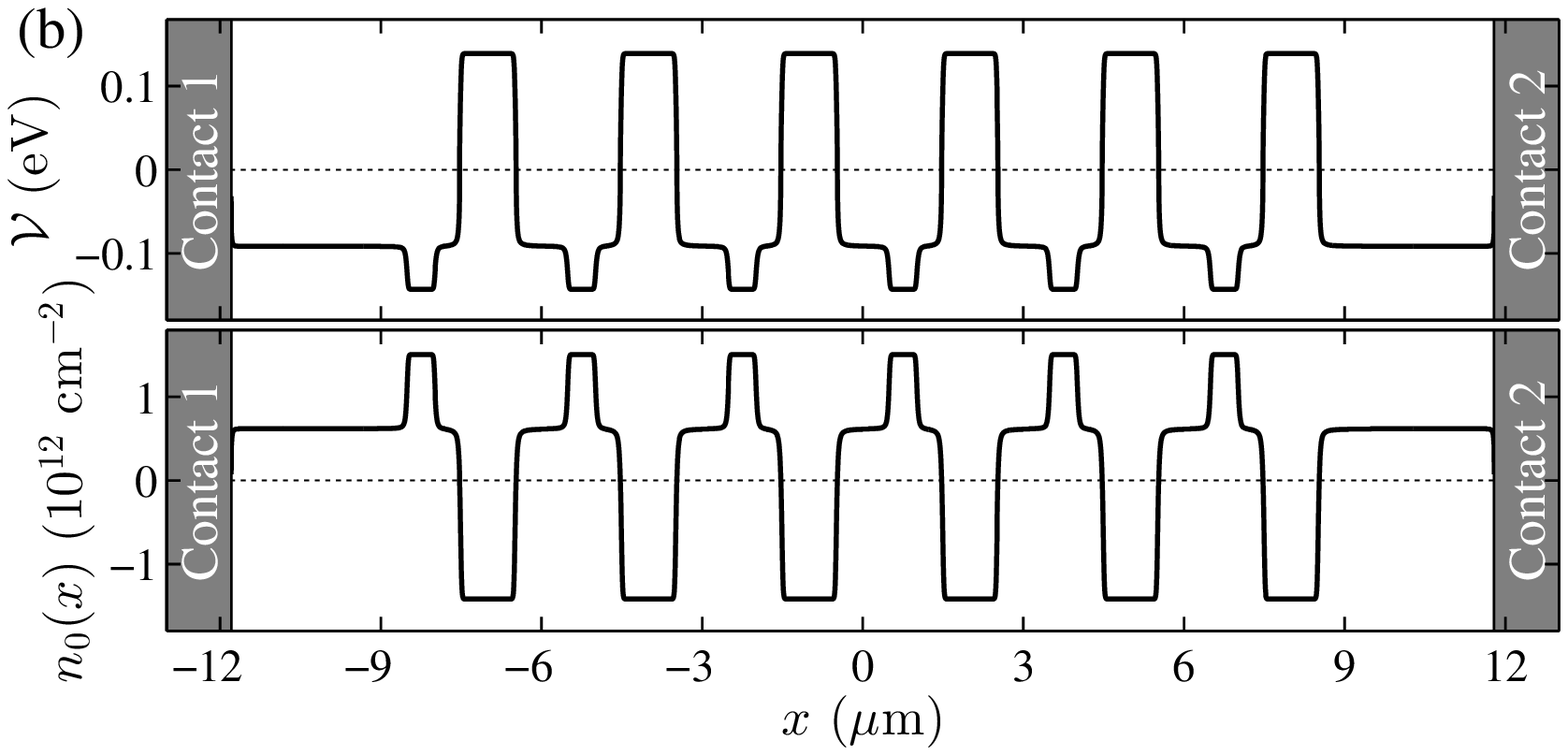}
\caption{
(a) Modeled DGG device (sample type \#B) geometry showing the spatial distribution of the electrostatic potential obtained 
by 2D finite-element simulation, considering gate voltages $(U_{\rm TG1},U_{\rm TG2})=(1,-2)\thinspace\mathrm{V}$ as an example. 
(b)~Extracted equilbrium free carrier density profile in graphene (lower panel, positive carrier density corresponds to electrons and negative carrier density corresponds to holes in graphene) 
and the corresponding local energy band offset (upper panel) assuming a uniform back gate contribution with voltage 
$U_{\rm BG}=10\thinspace\mathrm{V}$ as an example.
We use the energy units for the electrostatic potential in this paper.
}   
\label{fig_potential}
\end{figure}

The structures of the second type are fabricated on small area graphene flakes~\cite{Review0}. 
The benefit of these type of structures is the possibility to apply different bias voltages to the individual subgrating gates forming the superlattice allowing us to 
explore the role of 
the asymmetry of the lateral periodic electrostatic potential in the photocurrent formation 
as well as to examine the ratchet effects in the vicinity of the Dirac point. 
The graphene layers were prepared by mechanical exfoliation of natural graphite onto an oxidized 
silicon wafer. The samples used in this study were all single layer flakes. 
The periodic lateral electrostatic potential is created by {5 nm/60 nm  Ti/Au} 
inter-digitated metal-grating gates deposited on top of the graphene layer, 
see Fig.~\ref{fig_structure}(d)-(f), applying the method described above. 
The insulating layer of aluminum oxide is used to separate {the grating gates} and graphene. 
The asymmetric lateral structure incorporates  the
inter-digitated dual-grating gates (DGG)  TG1 and TG2 
having 
different stripe width and stripe separation.
An optical micrograph  of the interdigitated grating-gates is shown in 
Fig.~\ref{fig_structure}(e).
The supercell of the grating gate fingers consists of metal stripes
having two different widths $d_1 = 0.5~\mu$m and $d_2 = 1$~$\mu$m
separated by spacings $a_1 = 0.5$~$\mu$m and $a_2 = 1$~$\mu$m, 
Fig.~\ref{fig_structure}(d). This asymmetric supercell is 
 repeated six times to create a periodic asymmetric potential 
(period $d = 3$~$\mu$m), Fig.~\ref{fig_structure}(e). 
{The two subgrating gates, each formed by  fingers} of identical width,
can be biased independently.
Therefore, the asymmetry of the lateral potential {of the DGG structure} 
can be  varied in a controllable way. 
Figure~\ref{fig_potential} shows the potential 
profile obtained by a 2D
finite-element-based electrostatic simulation using
FENICS~\cite{FEniCS} and GMSH~\cite{gmsh}, for the device geometry of the experiment.
The profile of this potential was found by solving the Poisson 
equation taking into account its screening by the carriers 
in graphene and the quantum capacitance effect \cite{Luryi1988,Fang2007,Liu2013}; 
see Appendix~\ref{appendix esim} 
for details.
The samples were glued onto holders with conductive epoxy utilizing
the highly doped silicon wafer as a back gate
which enabled us to  change type and density of free carriers in graphene.

The experiments were performed using a $cw$ methanol laser~\cite{3aa,DMS2}
emitting at the frequency $f = 2.54$~THz  (wavelength of $\lambda$ = 118 $\mu$m and 
photon energy $\hbar\omega$ = 10.5 meV). 
The radiation power, $P$, being of the order of 50 mW at the sample surface, 
has been controlled by  pyroelectric detectors and focused 
onto samples by a parabolic mirror.
The  beam shape of the THz radiation is almost Gaussian, 
checked with a pyroelectric camera~\cite{Ganichev1999,Ziemann2000}. 
The peak intensity  in the laser spot on the sample, being of about 1.5~mm diameter~\cite{footnote}, 
was $I \approx 8$~W/cm$^2$. The THz radiation polarization 
state was varied by rotation of crystal quartz $\lambda$/4- and $\lambda$/2-plates~\cite{book}. 
To obtain circular and elliptically polarized radiation,
the quarter-wave plate was rotated by an angle $\varphi$ 
between the initial polarization plane and the optical axis of the plate. 
The radiation polarization states for several angles $\varphi$ are 
illustrated on top of Fig.~\ref{fig_pge}. 
The orientation of the linearly polarized radiation is defined by the azimuthal angle $\alpha$ 
with $\alpha\,=\,\varphi\,=\,0$  chosen in such a way that the electric 
field of the incident linearly polarized radiation is directed along  the $x$-direction, i.e. 
perpendicular to the metal fingers. 
The ratchet photocurrents have been measured in graphene structures at room 
temperature as a voltage drop across a 50 $\Omega$ or 470 $\Omega$ load resistance, $R_L$. 
The photovoltage  signal  is detected by using  standard lock-in technique.
The photocurrent ${\cal I}$ relates to the  photovoltage $V$ as ${\cal I}  =  V / R_L$ 
because in all experiments described below the load resistance was 
much smaller that the sample resistance $R_S$ ($R_L \ll R_S$). 
The corresponding photocurrent density is obtained as 
$j = {\cal I} / w$, where $w$ is the width  of the two-dimensional grating-gate structure 
being 1~mm for samples type \#A and 5~$\mu$m for samples type \#B.

\section{Experimental results}

\subsection{Photocurrents in large area epitaxial graphene structures}

\begin{figure}[t]
        \includegraphics[width=0.8\linewidth]{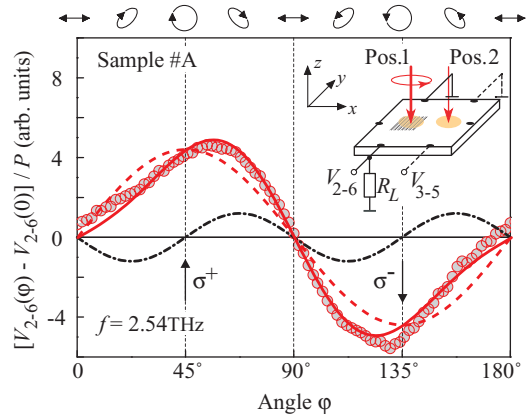}
\caption{Photovoltage $V_{2\mbox{-}6}(\varphi) -  V_{\rm off}$ measured in sample \#A for 
the laser spot focused on patterned graphene, position~1. The signal is plotted as a function of the angle 
$\varphi$ defining the radiation polarization state. The ellipses on top illustrate the polarization
states for several values of $\varphi$. 
Solid curve shows the fit according ${V_{2\mbox{-}6}(\varphi) -  V_{\rm off} = V_{\rm C} \sin 2\varphi  + 
V_{\rm L1}(\sin 4\varphi)/2}$, see also Eqs.~\eqref{s2},~\eqref{s3} and the corresponding discussion in the text. 
Dashed and dotted-dashed curves 
show individual contributions $V_{\rm C} \sin 2\varphi$ and $V_{\rm L1}(\sin 4\varphi)/2$, 
respectively. Arrows indicate angles $\varphi$ corresponding to right-handed 
($\sigma^+$) and left-handed ($\sigma^-$)
circularly polarized radiation.
}   
    \label{fig_pge}
\end{figure}

First we discuss the results obtained from the large scale lattice prepared 
on the top of  epitaxial graphene layer. Irradiating the structure 
with the THz beam, position~1 in Fig.~\ref{fig_pge}, we detected a polarization dependent 
photocurrent. Figure~\ref{fig_pge} shows the corresponding photovoltage, 
$V \propto j_y $, 
measured in the direction along the metal fingers ($y$-direction) as a function of 
the angle $\varphi$ governing the 
radiation ellipticity.
The signal varies with the radiation polarization 
as $V_{2\mbox{-}6}(\varphi) -  V_{\rm off} = V_{\rm C} \sin 2\varphi  + 
V_{\rm L1}(\sin 4\varphi) / 2$, where $V_{\rm off}$ is a polarization independent offset 
which is obtained as $V_{2\mbox{-}6}$ at $\varphi = 0$~\cite{footnote3}.
The figure reveals that the 
signal is dominated by the 
circular photocurrent $j_{\rm C} \propto V_{\rm C}$
being proportional to the degree of circular polarization 
$P_{\rm circ} = \sin 2\varphi $  and reversing its direction 
by  inverting of the THz radiation  helicity. 
The second contribution to the signal, $V_{\rm L1} \propto j_{\rm L1}$, whose amplitude is
about three times smaller than $V_{\rm C}$, 
corresponds to the photocurrent driven by the linearly polarized radiation and vanishes for 
circularly polarized radiation. 
In experiments applying half-wave plates it varies with the azimuth angle $\alpha$
as $V_{2\mbox{-}6}(\alpha) -  V_{\rm off} = V_{\rm L1}\sin 2\alpha$ (not shown). Note that 
the functions $(\sin 4\varphi)/ 2$ and $\sin 2\alpha$  describe the degree of linear polarization of THz electric field 
in the coordinate frame $x',y'$ rotated by $45^{\circ}$ in respect to the $x,y$ axes~\cite{Review_JETP_Lett}.
Our experiments were performed on several identically patterned samples. 
Though they were processed and structured identically, we observe different ratios of 
the linear and circular component for the same wavelength. 
As we show below, this fact can be attributed to slightly different 
intrinsic transport relaxation times (charge carrier mobilities) in the samples.

\begin{figure}[h]
        \includegraphics[width=0.8\linewidth]{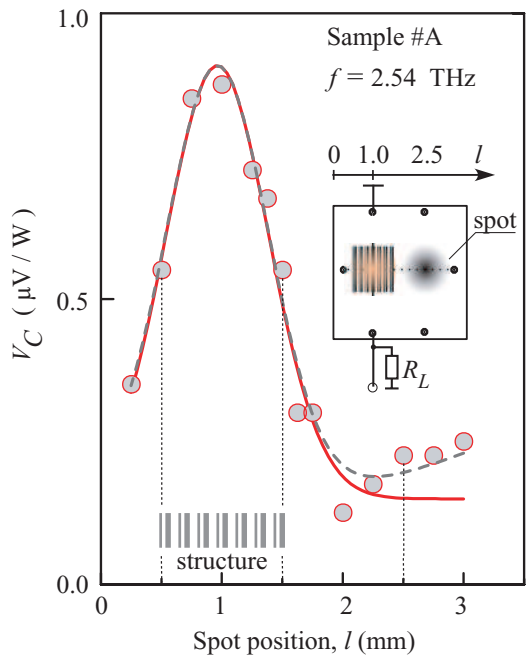}
\caption{Circular photocurrent $j_{\rm C} \propto V_{\rm C}$ in sample~\#A  as a function of the laser spot position.
The laser spot is scanned along the $x$-direction and the photocurrent is
picked up from contacts 2-6 allowing to probe current in the $y$-direction  (see top right inset).
The solid line represents the response calculated  assuming that the signal stems
from the lateral structure only and using the laser beam shape measured by the
pyroelectric camera. The curve is scaled to the photocurrent maximum. Dashed line is a guide for the eye.
The bottom inset shows schematically the grating-gate position. 
}   
\label{fig_scan}
\end{figure}

\begin{figure}[h]
        \includegraphics[width=0.9\linewidth]{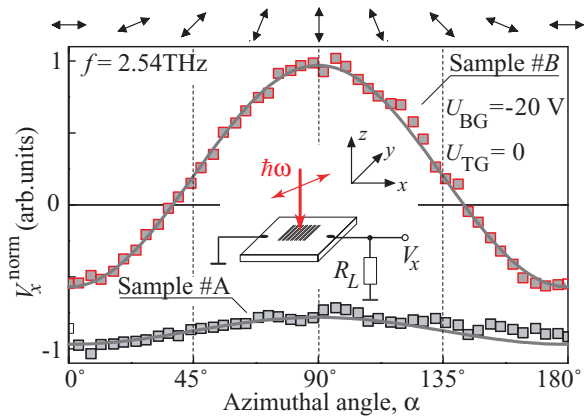}
\caption{Normalized photosignal $V^{\rm norm}_x$
induced by linearly polarized radiation  in the large epitaxial sample \#A and DGG graphene sample \#B  in the $x$-direction  normal to the metal stripes. Arrows on  top show the polarization plane orientation for several angles $\alpha$. 
The data for the DGG sample \#B are obtained for $U_{\rm TG1} = U_{\rm TG2} =0$ and $U_{\rm BG} = -20$~V. 
Solid curves show fit according to $V^{\rm norm}_{x} = V_0 + V_{\rm L2} \cos{2\alpha}$, see also Eq.~(\ref{s1}) and the corresponding discussion in the text.
}   
\label{fig_lp}
\end{figure}

Shifting the beam spot away from the structured area, 
position~2 in Fig.~\ref{fig_pge}, and measuring the signal either from contacts 2-6 or 3-5
we observe  the signal reduced by an order of magnitude.
This observation indicates that the photocurrent
stems from the irradiation of the superlattice. 
To provide additional evidence for this conclusion  we
scanned the laser spot across the sample along 
the $x$-direction. The photocurrent was measured 
between contacts 2 and 6 
aligned along the metal stripes, i.e. along the $y$-direction.
The experimental geometry and the circular photocurrent $j_{\rm C} \propto V_{\rm C}$ as a function
of the radiation spot position $l$  are shown in Fig.~\ref{fig_scan}.
The current reaches its maximum for the
laser spot centered at the superlattice
and rapidly decays with the spot moving away.
Comparison of ${V_{\rm C}(x)}$ with the 
curve calculated  assuming that the signal stems
from the lateral structure only and by using the laser beam 
spatial distribution measured by a pyroelectric camera
shows that the signal  follows this curve.
This observation unambiguously demonstrates that the
photocurrent is caused by irradiating the superlattice.
It also excludes photocurrents emerging due to 
 possible radiation-induced local heating causing the Seebeck 
effect, as such a signal should obviously reverse its sign 
at the middle of the sample.
The only deviation from this behaviour is detected for large 
values of $l$ at which the signal starts growing again. 
This result is attributed to the generation of 
the edge photocurrents reported in Refs.~\cite{Glazov2014,Karch_PRL_2011}.
For large $l$-values the beam spot reaches the edge of the graphene 
sample resulting in a photocurrent caused by the asymmetric 
scattering at the graphene edge~\cite{Karch_PRL_2011}. 
The ratchet photosignal $V_{1 \mbox{-}4} \propto j_x$ is also observed 
in the direction perpendicular to the fingers, i.e. along the $x$-direction.  
In this case the signal is insensitive to the THz electric field handedness and varies 
only with the degree of linear polarization as $V_{1\mbox{-}4} = V_0 + V_{\rm L2} {(\cos{4 \varphi}+1) / 2}$ or $V_{1\mbox{-}4} =  V_0 + V_{\rm L2} \cos{2\alpha}$, see Figure~\ref{fig_lp} showing $V_{1\mbox{-}4}$ 
as a function of the azimuthal  angle $\alpha$.
The same dependence has been measured in the DGG device, 
sample \#B in Fig.~\ref{fig_lp}, indicating that the DGG structure features 
the same superlattice effect as the large-area one (sample \#A). 
As we show below,  the appearance of the photocurrent along and accross the 
periodic structure as well as its polarization dependence are 
in full agreement with the ratchet effects excited by polarized THz electric field in asymmetric lateral superlattices. The overall 
qualitative behaviour of the photocurrent is also in agreement with that 
of the electronic ratchet effects observed in semiconductor quantum well structures
with a lateral superlattice~\cite{Olbrich_PRL_09,Olbrich_PRB_11,Review_JETP_Lett}. 
So far the  properties of graphene were not manifested explicitly.
The Dirac fermion properties of charge carriers in graphene manifest themselves
in superlattices  of type \#B with independently controlled gates.

\subsection{Photocurrents in inter-digitated dual-grating gates graphene structures}

\begin{figure}[h]
        \includegraphics[width=0.9\linewidth]{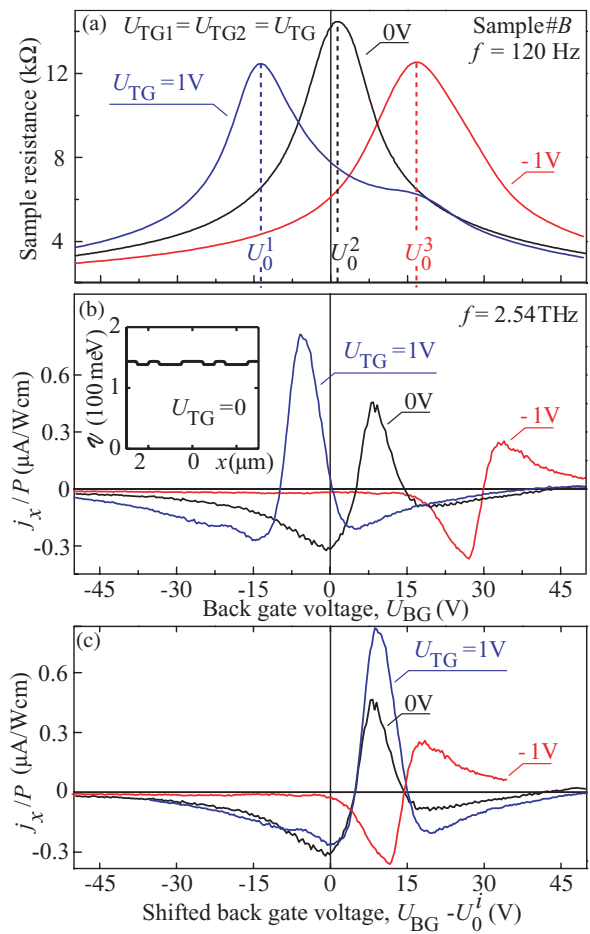}
\caption{
Back gate voltage dependencies obtained for the top subgate voltages $U_{\rm TG1} = U_{\rm TG2} = U_{\rm TG}$.
(a) Two terminal resistance of the DGG structure \#B. 
(b) Photocurrent $j_x(\alpha = 0)= j_0 + j_{\rm L2}$  normalized by the radiation intensity. Inset shows the 
energy band offset profiles at $U_{\rm TG}=0$ and $U_\mathrm{BG}=-20\thinspace\mathrm{V}$.
(c) Photocurrent $j_x(\alpha = 0)$ normalized by the radiation intensity as a function of the relative gate voltage $U_{\rm BG} - U^i_0$, 
where $U^i_0$ is defined as the back-gate voltage for which the resistance is the largest at corresponding $U_{\rm TG}$, see panel (a).
}
\label{fig_double1}
\end{figure}

The ratchet effects are expected to be strongly dependent
on the in-plane asymmetry of the electrostatic potential and near field 
effects of the radiation diffraction~\cite{Review_JETP_Lett,Theory_PRB_11,popov2013,Popov2015}. To demonstrate the effect of the asymmetry 
and examine 
the ratchet effects in the vicinity of Dirac point
we studied samples with an inter-digitated dual-grating structure, {see Figs.~1(d)-(f).} 
In the DGG structures the degree of asymmetry can be controllably varied by applying different potentials 
to the top grating gates. 
Moreover, using the back gate voltage $U_{\rm BG}$, we can globally change 
the background carrier density in the graphene flake. When the top gates are both grounded, 
the resistance exhibits one maximum upon tuning the back gate voltage.
The maximum corresponds to the Dirac point and is detected close to zero voltage, 
which confirms low residual doping of graphene, Fig.~\ref{fig_double1}(a). 
When we apply a voltage to the top gates, the back gate voltage corresponding to 
the Dirac point is shifted due to stray coupling of the top grating gates into the entire graphene. 
Furthermore, the top gates strongly modulate the carrier density in graphene directly 
underneath them, leading to an additional weak resistance maximum in the back gate response. 
Applying voltages of different polarity to different top grating gates results in a superpostition 
of resistance maxima corresponding to the Dirac points underneath and between the gate fingers 
(not shown here).
We observe a slight hysteresis in the back gate sweep (shift of the resistance 
maximum position by $U_{\rm BG} \approx 5$~V), 
which is probably due to the measurement being performed in the ambient air. 
When we set the TG1 voltage to $U_{\rm TG1} = -2$~V and TG2 voltage to $U_{\rm TG2} = +2$ V, 
we observe a splitting of the Dirac peak into three peaks (not shown), corresponding to regions with 
three different carrier densities:  underneath the top gates TG1 and TG2 
and in between the gate stripes, respectively.

Due to technological reasons (presence of the subgrating gates) the photocurrent in DGG structures
can be examined only in source-drain direction, i.e. normal to the gate stripes~\cite{footnote2}.
In the following experiments aimed to study the photocurrent $j_x(\alpha = 0)= j_0 + j_{\rm L2}$ 
as a function of the  back gate voltage for differently biased top gates  we used the
THz radiation polarized along the source-drain direction. 
Figure~\ref{fig_double1}(b) shows $j_x(U_{\rm BG})$ 
obtained for the three equal values of top gate voltages, $U_{\rm TG1} = U_{\rm TG2} = U_{\rm TG}$.
The photocurrent shows a complex sign-alternating behaviour with enhanced magnitude in the vicinity of the 
Dirac points being characterized by  resistance maxima and sign 
inversion for $U_{\rm TG} = 0 $ and  $U_{\rm TG} = +1$~V.
The photocurrents have opposite directions at very high (above $U_{\rm BG} = 40$~V) and at high negative back gate voltages.
Figure~\ref{fig_double1}(b) demonstrates that while the overall dependencies of
the  photovoltage obtained at different top gate potentials
are very similar they are shifted with respect to each other to that in 
 the transport curves. 
This 
is clearly seen in Fig.~\ref{fig_double1}(c) where
the curves for non-zero top gate voltages are 
shifted by the back gate voltage $U^i_{0}$
at which the resistance achieves maximum at corresponding $U_{\rm TG}$, 
see Fig.~\ref{fig_double1}(a). 
The figure reveals that the current can change its sign in the vicinity of the 
Dirac point. This fact can naturally be attributed to the change of 
carrier type from positively charged holes to negatively 
charged electrons~\cite{footnote4}.

\begin{figure}[h]
        \includegraphics[width=0.9\linewidth]{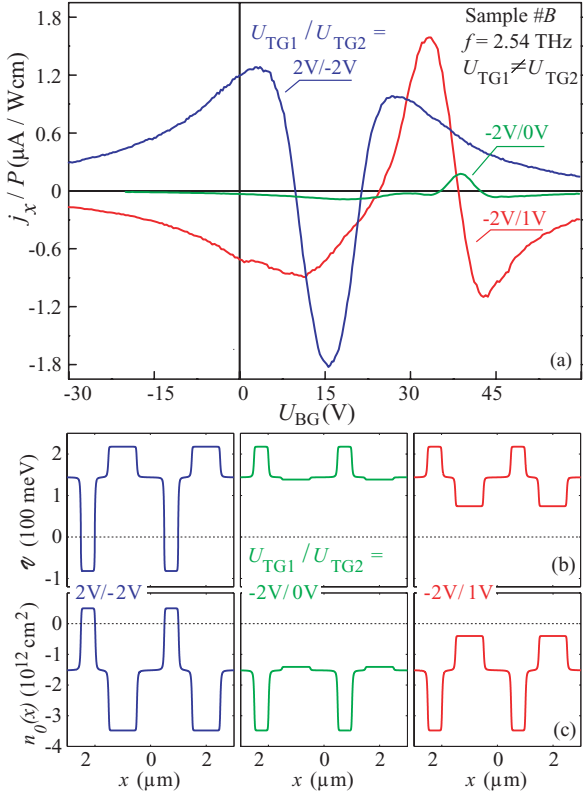}
\caption{  Gate voltage dependence of the photocurrent $j_x(\alpha = 0)= j_0 + j_{\rm L2}$ normalized by the radiation intensity measured for three sets of unequal potentials at the top gates, $U_{\rm TG1}\neq U_{\rm TG2}$. Insets show carrier density and energy band offset profiles at $U_{\rm BG}=-20\thinspace\mathrm{V}$.
}   
\label{fig_double2}
\end{figure}

In order to tune the lateral asymmetry, we applied different  bias voltages $U_{\rm TG1} \neq U_{\rm TG2}$ to the grating subgates.
Figure\,\ref{fig_double2} shows the photocurrent $j_x$ obtained for (i)~$U_{\rm TG1}= 2$~V, $U_{\rm TG2} = -2$~V, (ii)~$U_{\rm TG1}= -2$~V, $U_{\rm TG2} = 1$~V, and   (iii)~$U_{\rm TG1}= -2$~V, $U_{\rm TG2} = 0$.  For cases (i) and (ii), the potential asymmetry is efficiently inverted and hence we obtain 
inverted photocurrents far away from the Dirac point, i.e. at large values of $U_{\rm BG}$.
These observations show that  the photocurrent is caused by the excitation of the free carriers in graphene beneath the superlattice and 
its direction depends on the sign of the in-plane asymmetry of the electrostatic potential.
More complicated behaviour is detected in the vicinity of the  {Dirac points}. Here, the polarity of the free-carrier distribution in graphene and hence that of the photocurrent strongly depend  on the voltage set at the top gates. This  causes a more complicated variation of the photocurrent (including  its sign reversal) as a function of the back gate voltage, $U_{\rm BG}$,  around the Dirac point. Comparing the magnitudes of the signals for equal and unequal top gate voltages [Fig.\,\ref{fig_double1}(b) and Fig.\,\ref{fig_double2}(a), respectively] we see that in the latter case the photocurrent is several times enhanced. Finally, we note that if only one
of top gates is biased,  the signal  vanishes for almost all back gate voltages.
{The calculation of the electrostatic potential  indeed reveals that the 
potential becomes almost symmetric in this case indeed.} 

To summarize, experiments on two different types of graphene structures provide a self-consistent picture demonstrating that the photocurrents 
 (i) are generated due to the presence of asymmetric superlattices, (ii) are characterized by specific polarization dependencies for directions along and across the metal stripes, (iii) changes the direction upon reversing the in-plane asymmetry of the electrostatic potential as well as 
changing the carrier type,
(iv) are characterized by a complex sign-alternating back gate voltage dependence in the vicinity of the Dirac point, and (v) are strongly enhanced around the Dirac point.

\begin{figure}[h]
        \includegraphics[width=0.9\linewidth]{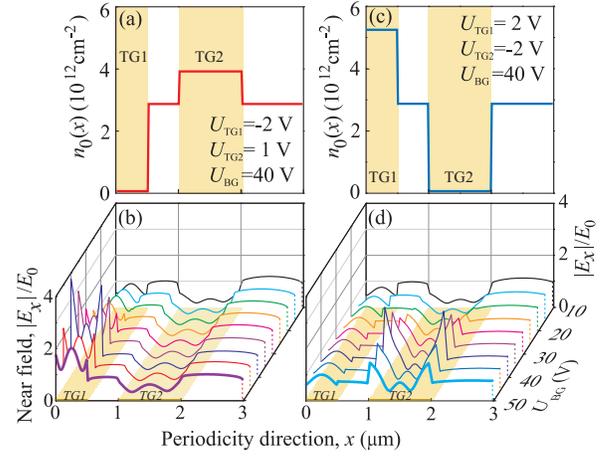}
\caption{Equilibrium free carrier density profile as a function of the coordinate $x$, shown for one period ($d= 3$~$\mu$m) of the DGG sample \#B. 
Shadowed areas indicate the positions of the metal stripes. The data are calculated for the electrostatic 
potential corresponding to: (a) $U_{TG1}=-2$~V, $U_{TG2}=1$~V, $U_{BG}=40$~V and (c) $U_{TG1}=2$~V, $U_{TG2}=-2$~V, $U_{BG}=40$~V. 
Panels (b) and (d) show the corresponding near-field amplitude distribution of the radiation with $f=2.54$~THz 
for the above set of gate potentials. 
The near-field is shown as a ratio of the amplitude of the near-field $|E_x|$  acting on graphene along the $x$-direction 
and the amplitude $E_0$ of the electric field of the incident plane wave.
}   
\label{nearfield}
\end{figure}

\section{Discussion}

Now we discuss the origin of the ratchet current in graphene with an asymmetrical grating irradiated by the THz beam. The effect of the grating is twofold: (i)~it generates a  one-dimensional periodic electrostatic potential ${\cal V}(x)$ acting upon the 2D carriers and (ii)~it causes a spatial modulation of the THz electric field due to the near field diffraction\,\cite{book}.
Figures~\ref{nearfield}(a) and~\ref{nearfield}(b) show  calculated coordinate dependencies of
 the free carrier density $n_0(x)$ and THz electric near-field $E(x)$ for the  DGG structure \#B
for two combinations of the top and back gate voltages. 
The electric field distribution caused by the near field diffraction is calculated for  radiation with the frequency $f=2.54$~THz applying  a self-consistent electromagnetic approach based on the integral equation method described in detail in Ref.~\cite{Fateev2010}, see also Appendix~\ref{appendix esim}.
Figures~\ref{nearfield} demonstrates that both the carrier density and THz field acting on charge carriers in the $x$-direction are asymmetric and
their distribution is strongly affected by the  voltages
applied to the individual top gratings.
These one-dimensional asymmetries result in the generation of a $dc$ electric current. As we show below the ratchet current may flow perpendicular to the metal fingers or along them. The mechanism leading to the photocurrent formation can be illustrated on the basis of the photocurrent caused by the Seebeck ratchet effect (thermoratchet). 
This type of the ratchet currents can be generated in the direction perpendicular to the metal stripes 
and corresponds to the photocurrent $j_x^S \propto V_0$ in Fig.~\ref{fig_lp}.
The spatially-modulated  electric field of the radiation heats the electron gas changing the effective electron temperature 
from the equilibrium value $T$ to $T(x) = \bar{T} + \delta T(x)$~\cite{footnoteLG}.
Here $\bar{T}$ is the average electron temperature and 
$\delta T(x)$ oscillates along the $x$-direction with the superlattice period 
$d$.
 In turn, the nonequilibrium correction $\delta T(x)$ causes an inhomogeneous
correction to the \textit{dc} conductivity,
$\delta \sigma(x) \propto \delta T(x)$. 
Taking into account the space modulated electric field $(-1/e) d{\cal V}/dx$ we obtain from Ohm's law 
the thermoratchet current in the form~\cite{Theory_PRB_11}
\begin{equation} \label{j_S}
j_x^S =- {1 \over e} \left< {d {\cal V} \over dx} \delta \sigma(x) \right>\:.
\end{equation}
Here $e<0$ is the electron charge, and angular brackets denote averaging over a spacial period. This photocurrent vanishes if the temperature is not space-modulated, therefore it is called the Seebeck ratchet current~\cite{footnoteT1}.

Besides the thermoratchet effect the THz radiation can induce additional photocurrents being
sensitive to the linear polarization plane orientation 
or to the helicity of circularly polarized photoexcitation. 
In the classical regime achievable in  our experiments and characterized by $\hbar \omega \ll \varepsilon_{\rm F}$ where $\varepsilon_{\rm F}$ is the Fermi energy, 
 all these photocurrents 
can be well described by means of 
Boltzmann's kinetic equation for the coordinate dependent distribution function $f_{\bm k}(x)$.
It has the following form:
\begin{equation} \label{Boltzmann}
\left( \frac{\partial}{\partial t} + v_{\bm k,x}
\frac{\partial}{\partial x} + \frac{{\bm F}(x)}{\hbar}
\frac{\partial}{\partial {\bm k}} \right)
f_{\bm k}(x) + Q_{\bm k} \{ f \}  = 0\:,
\end{equation} 
where ${\bm v}_{\bm k}$ is the velocity of an electron with the wavevector ${\bm k}$ being equal in graphene to $v_0 {\bm k}/k$, $v_0=10^6$~m/s is the Dirac fermion velocity,
 $Q_{\bm k}$ is the collision integral, and  ${\bm F}(x)$
is the  periodic force acting on charged carriers
\begin{equation} \label{force}
{\bm F}(x) = e \left[ {\bm E}(x) {\rm e}^{- {\rm i} \omega t} + {\rm c.c.} \right] - {d{\cal V}(x) \over dx} \, \hat{\bm x}\:,
\end{equation}
where $\hat{\bm x}$ is the unit vector in the $x$-direction.
In terms of the distribution function the electric current density is written as 
\begin{equation} \label{j}
{\bm j} = 4 e \sum\limits_{\bm k} {\bm v}_{\bm k} \left< {f_{\bm k}(x)} \right>\:,
\end{equation}
where the factor~4 accounts for the spin and valley degeneracies in graphene.
In the next section we present the theory of the ratchet currents which 
is valid for arbitrary large and abrupt periodic electrostatic potentials ${\cal V} (x)$,
and results of numerical calculations based on the developed theory and the complex distribution of the near field.
However, in order not to overload the discussion of the  experimental results
 with cumbersome equations, 
we first follow Refs.~\cite{Olbrich_PRL_09,Olbrich_PRB_11,Review_JETP_Lett,Theory_PRB_11}
and present solutions of the Boltzmann equation for 
weak and smooth electrostatic potential and the electric near-field.
Iterating the Boltzmann Eq.~\eqref{Boltzmann} for small ${\cal V}(x)$, $E(x)$ and their gradients, 
and ignoring the birefringence effect under the grating gate~\cite{IvchenkoPetrovFTT2014,footnoteStokes}, we obtain
the \textit{dc} current density components $j_x$ and $j_y$:
\begin{align} 
&j_x=\left<\left[ \chi_0 E^2 + \chi_L \, \left(|E_x|^2-|E_y|^2 \right) \right] {d{\cal V} \over dx} \right>, 
\label{phenom1}\\
&j_y=\left<\left[ \tilde{\chi}_L \, (E_xE_y^*+E_x^*E_y ) + \gamma \,{\rm i} (E_x E_y^* - E_x^* E_y) \right] {d{\cal V} \over dx} \right>.
\label{phenom2}
\end{align}
For incident radiation, the combinations bilinear in the field amplitudes  
vary upon rotation of quarter- and half-wave plates as~\cite{footnoteStokes} 
\begin{align}
\label{s1}
|E_x|^2-|E_y|^2 = E^2\frac{\cos{4\varphi} + 1}{2} = E^2\cos {2 \alpha},\\ 
\label{s2}
E_xE_y^*+E_x^*E_y= E^2\frac{\sin{4\varphi}}{2} = E^2\sin {2 \alpha}, \\ 
\label{s3}
{\rm i} (E_x E_y^* - E_x^* E_y) = 
-E^2\sin{2\varphi}.
\end{align}
All photocurrent contributions are detected in experiment, see 
Figs.~\ref{fig_pge} and \ref{fig_lp}. 
The coefficient $\chi_0$ corresponds to the thermoratchet  current discussed above. This photocurrent can be generated
in the in-plane direction normal to the metal stripes. 
 In experiments it yields the signal $V_0 \propto j_0$, see Fig.~\ref{fig_lp}.
The part of the signal detected for the same direction and varying upon rotation of the linear polarization, 
$V_{\rm LP2} \propto j_{\rm LP2}$ in Fig.~\ref{fig_lp}, is given by the second term in the right hand 
side of Eq.~(\ref{phenom1}) and is proportional to $\chi_L$~\cite{footnoteSeebeck}. 
The linear ($j_{\rm LP1} \propto V_{\rm LP1}  $) and circular  $ (j_{\rm C} \propto V_{\rm C})$ 
photocurrents observed in the direction along the metal stripes correspond to the first and second terms in 
the right hand side of Eq.~(\ref{phenom2}) and describe  
the linear ($\tilde{\chi}_L$) and circular ($\gamma$) ratchet effects, respectively.
The polarization dependent contributions appear because the free carriers in the laterally modulated graphene can move in the two directions ($x, y$) and are subjected to the action of the two-component electric field with the
$E_x$ and $E_y$ components.

Now we discuss the role of the superlattice asymmetry in the thermoratchet current formation.
Taking the lateral-potential modulation and the electric-field in the simplest form: 
\begin{equation}
{\cal V}(x) ={\cal V}_0 \cos{(qx + \varphi_{\cal V})},
\end{equation}
\begin{equation}
{\bm E}(x) = {\bm E}_0 [1 + h \cos{(qx + \varphi_E)}], 
\end{equation}
with $q = 2 \pi/d$,
we obtain for the average~\cite{Theory_PRB_11} 
\begin{equation} \label{E0V}
 \left< |\bm E(x)|^2 {d{\cal V} \over dx} \right> = q {\cal V}_0 h E_0^2 \sin{(\varphi_E-\varphi_{\cal V})}\:.
\end{equation}
The above phenomenological equations reveal that the thermoratchet current can be  generated only if the lateral superlattice is asymmetric. The asymmetry, created in our structures due to  different widths of the metal fingers and  spacings between them (Fig.~\ref{fig_structure}), causes a phase shift between the spatial modulation of the electrostatic lateral potential gradient $d{\cal V}(x)/dx$ and the near-field intensity $E^2(x)$ yielding a non-zero space average of their product.
The role of the superlattice lateral asymmetry and peculiarities of
the graphene band structure are explored in the experiments on the inter-digitated DGG structures. The back gate and the two independent top subgrating gates  enabled us to  controllably change the free carrier density profile in the $x$-direction.  
Let us begin with the data for  equal top gate potentials shown in 
Fig.~\ref{fig_double1}. At zero top gate voltages the asymmetry is created by the build-in potential 
caused by the metal stripes deposited on top of graphene. Transport data reveal that
at zero back gate voltage we deal with graphene almost at the charge neutrality point. 
The photocurrent shows a complex behaviour upon variation of the back gate voltage.
First of all it has the opposite polarities at large positive and negative back gate voltages. 
This fact can be primarily attributed to the change of the carrier type in graphene which 
results in the reversal of the current direction. At moderate back gate voltages 
the amplitude of the photocurrent substantially increases whereas in the vicinity 
of the Dirac point it exhibits a double sign inversion. The origin of this behaviour is unclear.
First of all, it may be caused by possible band-to-band optical transitions which become allowed
at  Dirac point  
because  the double Fermi energy can  be smaller than the photon energy in this case.
Also, as mentioned in the previous section,
the sign of the free-carrier distribution in graphene and hence that of the photocurrent strongly depend on the voltage set at the top gates which causes a more complex variation of the photocurrent (including a photocurrent sign reversal) as a function of the back gate voltage around the Dirac point.

Figure~\ref{fig_double1} demonstrates that the application of a positive or  negative voltage to both top gates 
does not change qualitatively the photocurrent behaviour but shifts the dependence as a whole 
to smaller or larger back gate voltages in  full correlation with the shift of the charge neutrality point
documented by transport measurements. These results show that the current is proportional to the charge sign of the carriers (negative for electrons, positive for holes) in graphene. Figure~\ref{fig_double1}(c)
reveals that  the photocurrent reverses its direction under inversion of the top gate voltage $U_{\rm TG}$ from +1~V to $-1$~V. This fact is also in agreement with Eq.~(\ref{phenom1}).
Indeed, at small amplitude of the potential, the photocurrent is $j \propto \left< |\bm E(x)|^2 d{\cal V}/dx\right>$.
This average changes sign upon the inversion of the potential 
${\cal V}(x) \to -{\cal V}(x)$. Even a more spectacular role of the in-plane asymmetry is exhibited in  the
experiments on the DGG structure with different polarities of the gate voltages applied to TG1 and TG2. 
First of all, the difference in the potentials increases the asymmetry resulting 
in the photovoltage enhancement  by more than an order of magnitude for large positive and negative gate voltages, Figs.~\ref{fig_double1}
and~\ref{fig_double2}. 
Moreover, the change of the relative polarity of the TG1 and TG2  gate voltages results in
a reversed photocurrent direction for all back gate voltages clearly reflecting the sign inversion of
the static potential asymmetry given by  $d{\cal V}/dx$.
Figure~\ref{fig_double2} also shows that for a certain combination of the top gate voltages
($U_{\rm TG1} = -2$~V and $U_{\rm TG2} = 0$) the photocurrent almost vanishes. This fact will be discussed in the next section presenting calculations of the photocurrent for exact profiles of the electrostatic potential and radiation near field.

Summarizing the discussion,  all experimental findings can consistently be explained qualitatively  by the free carrier ratchet effects.
A quantitative analysis 
is presented in the Sec.~\ref{theory_all}.

\section{Theory}
\label{theory_all}

\subsection{Photocurrent in the direction normal to the grating stripes}
\label{theory_1}

Three microscopic mechanisms of the ratchet current are considered and compared: 
(i)~the  Seebeck contribution generated in the course of 
the photoinduced spatially modulated heating of the free carriers accompanied by a periodic 
modulation of the equilibrium carrier density; 
(ii)~the polarization-sensitive current controlled by the elastic scattering processes and (iii) differential plasmonic drag. 
The main difference the first two mechanisms as compared with those in lateral quantum-well 
structures~\cite{Review_JETP_Lett} is determined by specific properties of graphene, namely, (i)~the linear, 
Dirac-like, dispersion of free-carrier energy, 
and (ii)~the degenerate statistics of the free-carrier gas in doped (or gated) graphene even at room temperature.

The photocurrent flowing in the periodicity direction is given by
\begin{equation}
    j_x = j_x^S
    + j_x^L + j_{\rm pl}^D.
\end{equation}
Here the first term is the Seebeck ratchet current.
The second term is caused by pure elastic scattering processes which 
are not related to carrier heating~\cite{Theory_PRB_11}, it yields a polarization dependent photocurrent 
varying upon rotation of the radiation polarization plane. 
The photocurrent $j_{\rm pl}^D$ is caused by the differential plasmonic drag~\cite{popov2013,Popov2015}.

We apply the kinetic theory for calculating the Seebeck ratchet current.
For degenerate statistics it yields (see Appendix~\ref{A2}):
\begin{equation}
\label{j_result}
    j_x^S =   {e^3 v_0^2 \over \pi \hbar^2} {\tau^2 \tau_\varepsilon \over 1 +(\omega\tau)^2} \frac{
    \left< |\bm E(x)|^2  {d \over dx} [\varepsilon_{\rm F}-{\cal V}(x)]^{-1} \right>
    }{ \left< [\varepsilon_{\rm F}-{\cal V}(x)]^{-1} \right>  }.
\end{equation}
Here $\tau$ and $\tau_\varepsilon$ are the free carrier momentum and energy relaxation times, respectively.
The derived expression for the Seebeck ratchet current is valid for arbitrarily large and abrupt periodic potential ${\cal V}(x)$.
We assume the Fermi energy to lie high above the Dirac point and take into account only one sort of free carriers, namely, the electrons. The similar results are obtained for the Fermi energy lying deep enough in the valence band in which case the electron representation is replaced by the hole representation.
Due to the charge-conjugation symmetry between electrons and holes in graphene, the current (\ref{j_result}) reverses under
the changes $\varepsilon_{\rm F} \to - \varepsilon_{\rm F}$, ${\cal V}(x) \to - {\cal V}(x)$ and ${e \to -e}$, where the energy is referred to the Dirac point.
We also note that this current vanishes if the $x$-coordinate dependence of the near field intensity $|\bm E(x)|^2$ is a composite
function $f[{\cal V}(x)]$. One more symmetry property follows for a low-amplitude potential $|{\cal V}(x)| \ll \varepsilon_{\rm F}$: in this case
the current reversal occurs just at the potential inversion ${\cal V}(x) \to - {\cal V}(x)$.

The differential plasmonic drag photocurrent $j_{\rm pl}^D$ induced in the grating-gated graphene by the normally incident THz radiation can be estimated 
as, see Appendix~\ref{A3},
\begin{equation}
j_{\rm{pl}}^D=- \frac{2 e^3 v_{0}^2}{\pi \hbar^2 \omega} 
\frac{\tau^2}{ 1 + (\tau \omega)^2} \sum_q q(|E_q|^2-|E_{-q}|^2),   
\label{5}
\end{equation}
where $E_q$ are the Fourier-space harmonics of the in-plane component of the near-electric field $E_x(x)$ in graphene with   $q=2\pi l/d$ where $l$ is an integer.
It is worth noting that the differential plasmonic drag~\cite{Popov2015} can be also viewed phenomenologically as a specific ''linear'' ratchet effect induced in a periodic graphene structure by the normally incident THz radiation
with the electric field polarized perpendicular to the grating gate.

We simulate the interaction of THz radiation incident normally upon the grating-gated graphene in the framework of a self-consistent electromagnetic approach based on the integral equation method described in detail in Ref.~\cite{Fateev2010}. 
The calculations are performed for the characteristic parameters of the DGG structure used in the experiment (see Appendix~\ref{appendix esim}).
In our simulations, we assume the metal grating stripes to be perfectly conductive and infinitely thin. This is a quite justified and commonly used assumption at THz (and lower) frequencies where metals are characterized by a high real conductivity. As a result of the electromagnetic modeling, we obtain the in-plane component of the near-electric field in graphene $\bm E(x) \parallel x$ entering Eqs.~(\ref{j_result}),~\eqref{5}. 
    Periodic electrostatic potential ${\cal V}(x)$ in graphene is created by applying different electric voltages to the two different subgratings of the DGG. 
    It should be noted that the periodic electrostatic potential is induced in graphene even for zero voltage at each subgrating due to finite density of states in graphene (the quantum capacitance effect). The profiles of the calculated near-electric field 
and the free-carrier density
    are shown in Figs.~\ref{nearfield} (b) and (d) for various voltages applied to different subgratings of the dual-grating gate at  frequency 2.54 THz. It is seen that the near-electric field is asymmetric relative to equilibrium free-carrier density profile in graphene. This gives rise to the Seebeck (thermoratchet) photocurrent Eq.~(\ref{j_result}).
    
\begin{figure}[h]
        \includegraphics[width=0.9\linewidth]{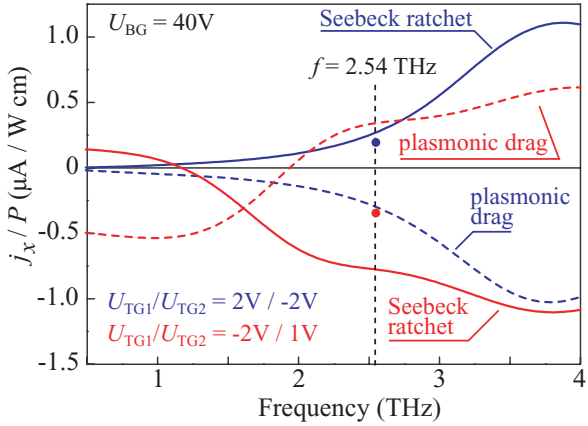}
\caption{Seebeck thermoratchet (solid lines) and plasmonic-drag (dashed line) photocurrents calculated for $U_{BG}=40$~V applied to the DDG graphene superlatice sketched in Fig.~\ref{fig_structure}. Red lines show the results for  $U_{TG1}$= 2V and $U_{TG2}$ = -2V, blue lines for $U_{TG1}$= -2V and $U_{TG2}$ = 1V. Red and blue dots correspond to the experimental data taken from Fig.~\ref{fig_double2}. 
}   
\label{fig_lambda_dep}
\end{figure}

    The calculated thermoratchet photocurrent as a function of frequency is shown by solid curves in Fig.~\ref{fig_lambda_dep} for various voltages applied to different subgratings of the DGG for monopolar graphene charged by applying 
large positive voltage to the back gate electrode. In this situation we deal with electrons in graphene under the metal fingers and between them even for negative voltages applied to a top gate.  It is worth noting that the magnitude of the photocurrents as well as the inversion of the photocurrent direction 
for reversing of relative signs  of voltages applied 
at different subgratings of the top dual-grating gate are in a accordance  with the experimental observations at the frequency 2.54~THz (Fig.~\ref{fig_lambda_dep}).

The calculated differential plasmonic drag photocurrent~(\ref{5}) as a function of frequency is shown by dashed curves in Fig.~\ref{fig_lambda_dep}. It follows from the figure that, as well as for the thermoratchet photocurrent, the inversion of the voltage signs at different subgratings of the top dual-grating gate changes the sign of the plasmonic-drag photocurrent.
However, the plasmonic-drag photocurrent is directed oppositely to the thermoratchet photocurrent.
Therefore, the plasmonic-drag photocurrent can compensate or, for a certain combination of top gate voltages,  even cancel the thermoratchet photocurrent diminishing the total photocurrent generated in graphene by the incident THz radiation. This fact may be responsible for the vanishingly small photocurrent observed for $U_{\rm TG1}= -2$~V and $U_{\rm TG2} = 0$,  Fig.~\ref{fig_double2}.

\subsection{Photocurrent in the direction along the grating stripes}
\label{theory_2}

Solution of the Boltzmann equation~\eqref{Boltzmann} also yields the $y$ component of the photocurrent.
In order to derive the expression for $j_y$ for the electrostatic potential ${\cal V}(x)$ comparable to the Fermi energy, we assume that both energy relaxation and diffusion are less effective than elastic scattering: $D_0(\pi/d)^2, \tau_\varepsilon^{-1} \ll \omega, \tau^{-1}$, but the product $\omega\tau$ can be arbitrary. 
Here $D_0=v_0^2\tau/2$ is the diffusion constant of the Dirac fermions in graphene.
We consider the elastic scattering by a long-range Coulomb potential. In this case, the momentum relaxation time in non-structured graphene is a linear function of the Fermi energy: $\tau\propto \varepsilon_{\rm F}$. Therefore, in the structures with a lateral superlattice under study we have ${\tau(x) = <\tau>\left[1-{\cal V}(x)/ \varepsilon_{\rm F}\right]}$. 
Using the procedure described in Ref.~\cite{Theory_PRB_11}, we obtain the photocurrent component along the grating stripes  in the form of Eq.~\eqref{phenom2}:
\begin{equation}
j_y=\left<\left[ \tilde{\chi}_L (E_xE_y^*+E_x^*E_y ) + \gamma {\rm i} (E_x E_y^* - E_x^* E_y) \right] {d{\cal V} \over dx} \right>,
\end{equation}
where the coordinate dependence of the coefficients $\tilde{\chi}_L$ and $\gamma$ is given by
\begin{align}
\label{gamma_chi}
&   \tilde{\chi}_L(x) = - {e^3 v_0^2 <\tau> \over 2\pi \hbar^2\varepsilon_{\rm F}} {\tau^2(x) [3+\omega^2\tau^2(x)] \over 1+\omega^2\tau^2(x)} , \\
&   \gamma(x) =  {e^3 v_0^2 <\tau> \over \pi \hbar^2\varepsilon_{\rm F}} {\tau(x)  \over \omega[1+\omega^2\tau^2(x)]}.
\nonumber
\end{align}
The coefficients $\tilde{\chi}_L$ and $\gamma$ describe the linear  and circular ratchet effects, respectively. 
The linear polarization-dependent and helicity-dependent combinations of the products $E_xE_y^*$ and $E_x^*E_y$ which determine the photocurrent $j_y$ are related to the corresponding combinations of the incident radiation as follows~\cite{IvchenkoPetrovFTT2014}:
\[ E_x(x)E_y^*(x)={\cal R}(x)E_{0x}E_{0y}^*.
\] 
A presence of imaginary part ${\rm Im}{\cal R}\neq 0$ is caused by effective birefringence of the studied low-symmetry structure resulting in an ellipticity of the near-field polarization under incidence of pure circularly or pure linearly (in the frame $x',y'$) polarized radiation.

\section{Conclusion}

To summarize, we have demonstrated that ratchet effects driven by THz electric fields
can be efficiently generated in graphene  with a lateral superlattice. 
The ratchet photocurrent includes the Seebeck thermoratchet effect 
as well as the effects of ``linear'' and ``circular'' ratchets, 
sensitive to the corresponding polarization of the driving electromagnetic force.
Studying the ratchet effect in the inter-digitated comb-like dual-grating-gate 
structures we have demonstrated that its amplitude and sign can be efficiently
controlled by applying unequal voltages to the DGG sublattices or 
the back gate voltage. 
We have calculated the electronic and plasmonic ratchet photocurrents at large negative and positive back
gate voltages taking into account the calculated 
potential profile and the near field acting on carriers in graphene.
The understanding of the observed complex back gate dependence and strong enhancement
of the ratchet effect in the vicinity of the Dirac point, however, 
requires further study. In particular, a theory describing ratchet effects for systems
with periodic change of the carrier type is to be developed. 
Further development of the theory is also required for a quantitative analysis 
of the plasmonic ratchet effects in graphene.

\appendix

\section{Electrostatic potential profile and near-field calculations}
\label{appendix esim}

The one-dimensional electric potential energy profile $\mathcal{V}(x)$ (or more precisely local energy band offset profile) is calculated via
\begin{align}
\mathcal{V}(x)=-\mathrm{sgn}[n_0(x)]\hbar v_0 \sqrt{\pi|n_0(x)|},
\label{V}
\end{align}
where $n_0(x)$ is the equilibrium carrier density profile (positive carrier density corresponds to electrons and negative carrier density corresponds to holes in graphene)
obtained by a 2D finite-element-based electrostatic simulation using \textsc{FEniCS} \cite{FEniCS} and \textsc{Gmsh} \cite{gmsh}, combined with the quantum capacitance model \cite{Fang2007,Liu2013}. 

The finite-element simulation follows the device geometry of the experiment, see Figs.~\ref{fig_structure} and~\ref{fig_potential}, and provides the classical self-partial capacitance for individual top gate set 1, $C_{\rm TG1}$, and set 2, $C_{\rm TG2}$, while the quantum capacitance model takes care of the correction to the net charge density due to the finite density of states of the conducting layer (here graphene)~\cite{Luryi1988}. Together with the global back gate capacitance $C_{\rm BG}$ that can be described by the parallel-plate formula, the total classical carrier density is given by
\begin{align}
n_C(x)=n_D+\sum_g\frac{C_g(x)}{|e|}U_g,
\label{nC}
\end{align}
where the summation runs over $g=\{{\rm TG1},{\rm TG2},{\rm BG}\}$, and a uniform intrinsic doping concentration $n_D\approx -9\times 10^{10}\thinspace\mathrm{cm}^{-2}$ is considered in our slightly \textit{p}-doped graphene sample. The net carrier density after taking into account the quantum capacitance correction reads \cite{Liu2013}
\begin{align}
\begin{split}
n_0(x)&=n_C(x)+\mathrm{sgn}[n_C(x)]n_Q(x)\left(1-\sqrt{1+2\frac{|n_C(x)|}{n_{Q}(x)}}\right) \\
&+\mathrm{sgn}(n_D)\sqrt{2n_Q(x)|n_D|},
\end{split}
\label{n}
\end{align}
with $n_C(x)$ given in Eq.~\eqref{nC} and $n_Q(x)=(\pi/2)[\hbar v_0/e^2\sum_g C_g(x)]^{2}$. Therefore Eq.~\eqref{n}gives the total carrier density as a function of position, subject to arbitrary voltage inputs, and can be inserted in Eq.~\eqref{V} to finally obtain the electric potential energy profile $\mathcal{V}(x\,;\,U_{\rm TG1},U_{\rm TG2},U_{\rm BG})$.

The near electric field in graphene induced by a normally incident THz wave was calculated by using the self-consistent electromagnetic approach described in Ref.~\cite{Fateev2010}. Calculations were performed for the characteristic parameters of the DGG structure used in the experiment: $d_1=0.5~\mu\rm m$, $d_2=1~\mu\rm m$,  $a_1=0.5~\mu\rm m$,  $a_2=1~\mu\rm m$, and $\tau=5~\rm{ps}$. Dielectric constants of the graphene substrate ($\rm{SiO_2}$) and the barrier layer ($\rm{Al_2O_3}$) between graphene and the top DGG gate are 3.9  and 9 (see~\cite{epsilon}), respectively. 
The barrier layer thickness is 30~nm.
The frequency-dependent response of graphene is described by a local dynamic conductivity~\cite{Falkovsky2007}
\begin{align} \label{Popov1}
& \sigma(\omega)= \sigma_0 \Biggl\{\frac{8k_{\rm B}T\tau}{\pi\hbar\left(1-i\omega\tau\right)}\ln \left[ 2 \cosh \left( \frac{\varepsilon_{\rm F}-{\cal V}(x)}{2 k_{\rm B}T}\right) \right] \nonumber \\
& +G\left( \frac{\hbar \omega}{2}\right)+\frac{4i\hbar\omega}{\pi}\int\limits_{0}^{\infty} \frac{G\left(\varsigma\right)-G\left( \frac{\hbar \omega}{2}\right)}{(\hbar \omega)^2-4\varsigma^2}d\varsigma \Biggr\},
\end{align}
where
\begin{equation} \label{Popov2}
G(\varsigma)=\frac{\sinh(\varsigma/k_{\rm B}T)}{\cosh(\varepsilon_{\rm F}/k_{\rm B}T)+\cosh(\varsigma/k_{\rm B}T)},
\end{equation}
$\sigma_0=e^2/(4\hbar)$, 
and the temperature $T$ is set to 300\,K.
The first term in Eq.~(\ref{Popov1}) describes a Drude-like response involving the intraband processes with the phenomenological carrier scattering time $\tau$, which can be estimated from the measured $dc$ carrier mobility as $\tau=\mu \varepsilon_{\rm F}/(|e|v_0^2)$~\cite{Tan2007}.   The second and third terms in Eq.~(\ref{Popov1}) describe the interband carrier transitions in graphene.

\section{Derivation of the Seebeck ratchet current density}
\label{A2}

For the structures under consideration one needs to extend the theory of the Seebeck ratchet current derived in Ref.~\cite{Theory_PRB_11} for a weak electron periodic potential ${\cal{V}}(x)$ to an arbitrarily large potential. For simplicity, we assume the Fermi energy to lie high enough above the Dirac point 
and consider one sort of free carriers. 

The absorption of THz radiation results in an inhomogeneous heating of 2D carriers in graphene with a lateral superlattice. Similarly to Ref.~\cite{Theory_PRB_11}, we present the time-independent electron distribution function $f_{\bm k}$ as a sum  $f^+_{\bm k} + f^-_{\bm k}$ of the components even and odd in ${\bm k}$, respectively, decompose the Boltzmann kinetic equation into even and odd parts, select the odd-in-${\bm k}$ part and arrive at the following equation for the function $f^-_{\bm k}$ contributing to the Seebeck ratchet effect 
\begin{equation} \label{f-}
{f_{\bm k}^- \over \tau} + v_x{\partial f_{\bm k}^+ \over \partial x} - {1\over \hbar}{\partial V \over \partial x}{\partial f_{\bm k}^+ \over \partial k_x} = 0 .
\end{equation}
Here the potential $V(x)$ is a sum ${\cal V}(x) + U(x)$ of the equilibrium potential ${\cal V}(x)$ and a correction $U(x)$ that appears due to the radiation-induced carrier redistribution. This correction is proportional to the radiation intensity and related by the Poisson equation to a radiation-induced change $\delta n(x) = n(x)-n_0(x)$ where $n_0(x)$ is the equilibrium electron density and \begin{equation} \label{nf+}
n(x) = 4 \sum_{\bm k} f_{\bm k}^+(x)
\end{equation} 
is the steady-state nonequilibium density. The Poisson equation is easier to express in terms of the Fourier transforms as follows
\begin{equation} \label{Poisson}
2 |q_x| \mbox{\ae} U_{q_x} = 4 \pi e^2 \left( {n}_{q_x} - n_{0,q_x} \right)\:,
\end{equation}
where $\mbox{\ae}$ is the dielectric constant. Hereafter the average value of the electron density $\bar{n}$ is fixed which imposes the following restriction on the electron density 
\begin{equation} \label{restr}
\frac{1}{d}\int\limits_0^d n(x) dx = \frac{1}{d}\int\limits_0^d n_0(x) dx \equiv \bar{n}\:.
\end{equation}

Multiplying the equation (\ref{f-}) by $e v_{{\bm k}x} \tau$ and summing over the electron spin $s$, the valley index $\nu$ 
and the quasi-momentum $\bm k$ we get the Seebeck ratchet current density:
\begin{equation} \label{jSeebeck}
    j = e{d V \over dx} \sum_{\nu s \bm k} \tau v_x^2 {\partial f_{\bm k}^+\over \partial \varepsilon_k} - e \sum_{\nu s \bm k} \tau v_x^2 {\partial f_{\bm k}^+ \over \partial x} \:,
\end{equation} 
where $\varepsilon_k$ is the electron energy dispersion linear in graphene.
The current is zero in the absence of radiation because, in equilibrium, $f_{\bm k}^+(x)=f_0[\varepsilon_k + {\cal V}(x)]$ where $f_0(\varepsilon_k)$ is the Fermi-Dirac distribution function, and hence 
\[
{\partial f_{\bm k}^+(x) \over \partial x} = {d {\cal V}(x) \over dx} {\partial f_{\bm k}^+(x) \over \partial \varepsilon_k}\:.
\]

The current density (\ref{jSeebeck}) can be expressed via the conductivity 
\begin{equation} \label{sigmaf+}
\sigma = e^2  \sum_{\nu s \bm k} {v_k^2 \tau \over 2}\left(-{\partial f_{\bm k}^+ \over \partial \varepsilon_k} \right)
\end{equation}
and the diffusion coefficient $D(\varepsilon_k) = v_k^2\tau/2$ as follows
\begin{equation} \label{j0}
    j = -{1 \over e} {d V \over dx} \sigma(x) - e {\partial \over \partial x}\sum_{\nu s\bm k} D(\varepsilon_k) f_{\bm k}^+ .
\end{equation}
This result is valid in all orders in ${\cal V}(x)$. 

In what follows we assume $\tau$ to be independent of the 
particle energy $\varepsilon_k$. Then in graphene the coefficient $D$ equals to $v_0^2\tau/2 \equiv D_0$ and is independent of $\varepsilon_k$. As a result, one has
\begin{equation}
j = - {1 \over e}{d V(x) \over dx} \sigma(x)    -eD_0 {d n(x) \over d x} .
\end{equation}
Obviously, one can equivalently substitute the corrections $\delta n(x)$ and $\delta \sigma(x)$ into Eq.~(\ref{j0}) instead of
$n(x)$ and $\sigma(x)$. 

Before the search for $\delta n(x)$ and $\delta \sigma(x)$ we exclude the potential $U(x)$ from the consideration. For this purpose we decompose the electron density and the conductivity in the following form
\begin{equation} \label{first}
n(x) = \tilde{n}_0(x) + \delta n'(x)\:,
\quad
\sigma(x) = \tilde{\sigma}_0(x) + \delta \sigma'(x) + \delta\sigma_{\rm heat}\:.
\end{equation}
Here $\delta \sigma_{\rm heat}(x)$ is a local change of the electron conductivity caused by heating by the THz radiation, see below,
and the functions $\tilde{n}_0(x)$ and $\tilde{\sigma}_0$ are auxiliary: they are found from Eqs. (\ref{nf+}) and (\ref{sigmaf+}) with the exact function $f_{\bm k}^+$ replaced by the auxiliary 
(quasi-equilibrium) function
\begin{equation} \label{fplus}
\tilde{f}^+_k = \left\{ \exp{\left[ \varepsilon_k - \varepsilon_{\rm F}-\delta\varepsilon_{\rm F} + {\cal V}(x) + U(x)\over T \right]} +1 \right\}^{-1}\:,
\end{equation}
where $\varepsilon_F$ is the equilibrium Fermi energy and the correction $\delta \varepsilon_F$
restores the average electron density. One can check that $\tilde{n}_0(x)$ and $\tilde{\sigma}_0(x)$ satisfy the equation
\begin{equation} 
- {1 \over e} \tilde{\sigma}_0(x) \frac{d}{dx} \left[ {\cal V}(x) + U(x) \right] -eD_0 {d\tilde{n}_0(x) \over d x}= 0\:.
\end{equation}
Neglecting a second-order correction proportional to $[dU(x)/dx] \delta \sigma'(x)$ we obtain an equation for the electric current determined exclusively by ${\cal V}(x)$ and the corrections $\delta n'(x)$, $\delta \sigma'(x)$:
\begin{equation}
\label{j1}
    j = - {1 \over e}{d {\cal V} \over dx} [\delta \sigma'(x)+\delta\sigma_{\rm heat}(x)]   -eD_0 {d  \delta n' \over d x}\:.
\end{equation}
The two corrections are related with each other by 
\begin{equation}
    \delta \sigma'(x) = g_0(x)\delta n'(x), \quad g_0(x)={\delta \sigma_0(x) \over \delta n_0(x)}\Biggr|_{T={\rm const}}.
\end{equation}
Here we take into account that the correction $\delta n'(x)$ is caused by redistribution of carriers but not by heating. In contrast, the correction $\delta \sigma_{\rm heat}(x)$ is due to heating at a fixed carrier density:
\begin{equation}
    \delta \sigma_{\rm heat}(x) = h_0(x)\delta \varepsilon_{\rm heat}(x), \quad h_0(x)={\delta \sigma_0(x) \over \delta \varepsilon_0(x)}\Biggr|_{n={\rm const}}\:,
\end{equation}
where $\varepsilon_0(x)$ is the average electron energy in equilibrium
and $\delta \varepsilon_{\rm heat}(x)$ is a  local change of the electron average energy caused by the THz radiation. As above, the index ``0'' denotes functions calculated in the absence of radiation and dependent on the coordinate $x$ due to the $x$-dependence of the static potential ${\cal V}$.
The change $\delta \varepsilon_{\rm heat}(x)$ is found from the energy balance equation
\begin{equation} \label{heating}
{n_0(x) \delta \varepsilon_{\rm heat}(x) \over \tau_\varepsilon} = 2|E(x)|^2 {\sigma_0(x) \over 1 +(\omega\tau)^2}\:, 
\end{equation}
where $\tau_{\varepsilon}$ is the energy relaxation time. Thus, Eq.~(\ref{j1}) reduces to the equation
\begin{equation}
\label{diff_ur_gr}
    j = - {1 \over e}{d {\cal V}(x) \over dx}  \left[g_0(x)\delta n'(x) + h_0(x)\delta \varepsilon_{\rm heat}(x) \right]    -eD_0 {d \delta n'(x) \over d x}
\end{equation}
containing one unknown function, $\delta n'(x)$.

Under the requirement of coordinate independence of $j$, Eq.~\eqref{diff_ur_gr} forms a first-order differential equation for the correction $\delta n'(x)$. Its solution is given by
\begin{align} \label{deltanprime}
&   \delta n'(x) = {\rm e}^{-F_0(x)} \Biggl\{ \delta n'(-d/2)  \\
     & - {1 \over eD_0} \int\limits_{-d/2}^x dx' {\rm e}^{F_0(x')} 
    \left[ {1 \over e}{d {\cal V} \over dx'} h_0(x')\delta \varepsilon_{\rm heat}(x') + j \right]
    \Biggr\}\:, \nonumber
\end{align}
where
\begin{equation} \label{Fx} 
    F_0(x) =  {1 \over e^2 D_0}\int\limits_{-d/2}^x dx_1 {d {\cal V}(x_1)  \over dx_1} g_0(x_1),
\end{equation}
and a value of $\delta n'(-d/2)$ is determined from the condition (\ref{restr}).
Moreover, after averaging over the space period in Eq.~(\ref{diff_ur_gr}) the third contribution vanishes and we obtain
\begin{equation} \label{javerage}
    j = - {1 \over e} \left<
    {d {\cal V} \over dx} \left[g_0(x) \delta n'(x) + h_0(x)\delta \varepsilon_{\rm heat}(x)\right]
    \right>,
\end{equation}
where the angular brackets mean averaging over $x$. According to Eq.~(\ref{deltanprime}) the correction $\delta n'(x)$ depends linearly on $j$. Therefore, Eq.~(\ref{javerage}) is just a linear algebraic equation for the Seebeck ratchet photocurrent. Substituting $\delta n'(x)$ from Eq.~(\ref{deltanprime}) and noticing that the term with $\delta n'(-d/2)$ does not contribute to the current [because the function $F_0(x)$ depends on the coordinate $x$ via the static potential ${\cal V}(x)$] we finally obtain
\begin{align}
\label{j_gr}
    j =  -{1 \over e{(1-\alpha)} } \Biggl<
    \delta \varepsilon_{\rm heat}(x) {d {\cal V} \over dx}h_0(x) \\
    \times \left[1 - {\rm e}^{F_0(x)}\int\limits^{d/2}_x dx' {\rm e}^{-F_0(x')} 
     {d {\cal V} \over dx'} {g_0(x') \over e^2D_0} \right]
    \Biggr>\:, \nonumber
\end{align}
where
\begin{equation}
    \alpha = {2 \over e^2 D_0} \int\limits_{0}^{d/2} \frac{dx'}{d} {\rm e}^{F_0[{\cal V}(x')]}  \int\limits^{{\cal V}(d/2)}_{{\cal V}(x')} dV  g_0(V) {\rm e}^{-F_0(V)} .
\end{equation}

Now we calculate the variational derivatives $g_0(x)$ and $h_0(x)$. The even part of the distribution function modified by the temperature change $\delta T(x)$ is given by
\begin{equation}
    f^+_k = \left\{ \exp{\left[ \varepsilon_k - \varepsilon_{\rm F}-\delta\varepsilon_{\rm F}(x) +{\cal V}(x) \over T + \delta T(x) \right]} +1 \right\}^{-1},
\end{equation}
where $\delta T(x)$ is a radiation-induced change of the local electron temperature and, in contrast to Eq.~(\ref{fplus}), the correction $\delta\varepsilon_{\rm F}(x)$ is varying in space due to the electron redistribution following the inhomogeneous heating. 
The functions $\delta \sigma_0(x)$, $\delta \varepsilon_0(x)$ and $\delta n_0(x)$ are expressed via $\delta\varepsilon_{\rm F}(x)$ and $\delta T(x)$ 
\begin{eqnarray}
&&\delta \sigma_0(x) = {\partial \sigma_0 \over \partial \varepsilon_{\rm F}} \delta\varepsilon_{\rm F}(x)
    + {\partial \sigma_0 \over \partial T} \delta T(x)\:, \\
&&\delta \varepsilon_0(x) = {\delta \varepsilon_0 \over \delta \varepsilon_{\rm F}(x)} \delta\varepsilon_{\rm F}(x)
    + {\delta \varepsilon_0 \over \delta T(x)} \delta T(x)\:,\\
&&\delta n_0(x) = {\delta n_0 \over \delta \varepsilon_{\rm F}} \delta\varepsilon_{\rm F}(x)
    + {\delta n_0 \over \delta T(x)} \delta T(x)\:,
    \end{eqnarray}
from whence we obtain
\begin{equation} \label{gxhx}
    g_0(x) = {\partial \sigma_0 / \partial \varepsilon_{\rm F}\over \partial n_0 / \partial \varepsilon_{\rm F}},
    \qquad h_0(x) = \frac{\Delta^-_{\sigma n}}{\Delta^-_{\varepsilon n}}\:,
\end{equation}
\[
\Delta^-_{\sigma n} = {\partial \sigma_0 \over \partial \varepsilon_{\rm F}}{\partial n_0 \over \partial T} -
    {\partial \sigma_0 \over \partial T}{\partial n_0 \over \partial \varepsilon_{\rm F}}\:,
    \quad
\Delta^-_{\varepsilon n} =  {\partial \varepsilon_0 \over \partial \varepsilon_{\rm F}}{\partial n_0 \over \partial T} - {\partial \varepsilon_0 \over \partial T}{\partial n_0 \over \partial \varepsilon_{\rm F}} \:.
\]

In equilibrium, the concentration, average particle energy and conductivity are obtained from the corresponding values in unstructured graphene, $n_0(\varepsilon_{\rm F})$, $\varepsilon_0(\varepsilon_{\rm F})$ and $\sigma_0(\varepsilon_{\rm F})$, by the substitution
\begin{equation}
\varepsilon_{\rm F} \to \varepsilon_{\rm F}-{\cal V}(x).
\end{equation}
These values depend on the Fermi energy and temperature as follows ($k_{\rm B} \equiv 1$, spin and valley degeneracies are taken into account):
\begin{equation}
    n_0  
     = {T^2 \over \pi (\hbar v_0)^2} \left[ {\varepsilon_{\rm F}^2 \over T^2} + {\pi^2 \over 3} + 2{\rm Li}_2 \left(-{\rm e}^{-{\varepsilon_{\rm F}\over T}}  \right)\right],
\end{equation}
\begin{equation}
    n_0 \varepsilon_0 
     = {2T^3 \over 3\pi (\hbar v_0)^2} \left[ {\varepsilon_{\rm F}^3 \over T^3} + {\pi^2}{\varepsilon_{\rm F} \over T} + 3{\rm Li}_3 \left(-{\rm e}^{-{\varepsilon_{\rm F}\over T}}  \right)\right],
\end{equation}
\begin{equation}
    \sigma_0 
    = {2T e^2 D_0 \over \pi (\hbar v_0)^2} \ln{\left( 1 + {\rm e}^{\varepsilon_{\rm F}\over T} \right) }.
\end{equation}
Here ${\rm Li}_{2,3}(z)$ are the polylogarithm functions of orders 2 and 3, respectively.
At $T \ll \varepsilon_{\rm F}$ these expressions reduce with accuracy up to $(T/\varepsilon_{\rm F})^2$:
\begin{align}
\label{T02}
&   n_0 = {1\over \pi (\hbar v_0)^2} \left(\varepsilon_{\rm F}^2 + {\pi^2 \over 3}T^2\right),
\\
&\varepsilon_0 =  {2\over 3} \varepsilon_{\rm F}+ {4\pi^2 \over 9}{T^2 \over \varepsilon_{\rm F}},  
\quad
\sigma_0 =  {2 \varepsilon_{\rm F} e^2 D_0 \over \pi (\hbar v_0)^2}.
\end{align}
This allows us to calculate the functions $g_0(x)$ and $h_0(x)$:
\begin{equation}
    g_0(x) = {e^2 D_0 \over \varepsilon_{\rm F}-{\cal V}(x)},
    \qquad
    h_0(x) = -{e^2 D_0 \over \pi (\hbar v_0)^2}.
\end{equation}
Then we obtain:
\begin{equation}
    F_0(x) 
    = \ln{\left[ {\varepsilon_{\rm F}-{\cal V}(d/2) \over \varepsilon_{\rm F}-{\cal V}(x)}\right]},
\end{equation}
\begin{equation}
    \alpha 
    = 1- \left< {\varepsilon_{\rm F}-{\cal V}(d/2) \over \varepsilon_{\rm F}-{\cal V}(x)}\right>,
\end{equation}
and proceed from Eq.~\eqref{j_gr} for the ratchet current to
\begin{equation}
    j =  {e \tau \over 2\pi \hbar^2} \left<
    \delta \varepsilon_{\rm heat}(x) {d {\cal V} \over dx} {1\over \varepsilon_{\rm F}-{\cal V}(x) } \right>
      \left< {1 \over \varepsilon_{\rm F}-{\cal V}(x)}\right>^{-1}.
\end{equation}

The energy balance equation (\ref{heating}) yields
\begin{equation}
    \delta \varepsilon_{\rm heat}(x) = 2|E|^2 {e^2 v_0^2 \tau \over 1 +(\omega\tau)^2} {1 \over \varepsilon_{\rm F}-{\cal V}(x)}.
\end{equation}
    Therefore we finally arrive at
\begin{equation}
\label{j_gr_gen}
    j =   {e^3 v_0^2 \over \pi \hbar^2} {\tau^2 \tau_\varepsilon \over 1 +(\omega\tau)^2} {
    \left< |E(x)|^2  {d\over dx} [\varepsilon_{\rm F}-{\cal V}(x)]^{-1} \right> \over
     \left< [\varepsilon_{\rm F}-{\cal V}(x)]^{-1} \right> }.
\end{equation}

For ratchets based on quantum-well structures with a parabolic energy dispersion $\varepsilon_{\bm k}=\hbar^2k^2/(2m)$, the analogous procedure yields the Seebeck ratchet current density in the form:
\begin{align}
\label{j_gen1}
&   j =   {\bar{n}e\tau \over m \varepsilon_{\rm F}} 
    \left< {\varepsilon_{\rm F} -{\cal V}(d/2)\over \varepsilon_{\rm F}-{\cal V}(x)} \right>^{-1}  
    \Biggl< \delta \varepsilon_{\rm heat}(x) \\ \times
& {d\over dx} 
    \left\{ 3{\cal V}(x) + 2[\varepsilon_{\rm F}- {\cal V}(d/2)] \ln{\left[ \varepsilon_{\rm F}- {\cal V}(d/2) \over \varepsilon_{\rm F}-{\cal V}(x)\right]}  \right\}\Biggr>. \nonumber
\end{align}
In this case, $\delta \varepsilon_{\rm heat}(x)$ is given by
\begin{equation}
    \delta \varepsilon_{\rm heat}(x) = 2|E(x)|^2 {e^2 \tau/m \over 1 +(\omega\tau)^2}.
\end{equation}

\section{Differential  plasmonic drag in graphene}
\label{A3}

Let us consider a homogeneous graphene screened by an inter-digitated metal DGG. We simulate the plasmonic response of graphene by solving the hydrodynamic equations
\begin{equation}
e \frac{\partial n(x,t)}{\partial t}+\frac{\partial j(x,t)}{\partial x}=0 \,\, ,
\label{A4_1}
\end{equation}
\begin{multline}
\frac{\partial v(x,t)}{\partial t}+v(x,t)\frac{\partial v(x,t)}{\partial x}+\frac{v(x,t)}{\epsilon_{\rm F}(x,t)} \frac{\partial \epsilon_{\rm F}(x,t)}{\partial t} \\
=\frac{e v_{\rm 0}^2}{\epsilon_{\rm F}(x,t)} E(x,t)-\frac{v(x,t)}{\tau},
\label{A4_2}
\end{multline}
describing the free-carrier motion in graphene, where $j(x,t)=en(x,t)v(x,t)$ is the electric (in general, nonlinear) current in graphene, $n(x,t)$ and $v(x,t)$ are the charge density and hydrodynamic velocity of the carriers in graphene, respectively, $\epsilon_{\rm F}(x,t)$ is the Fermi energy in graphene related to the carrier density as $\epsilon_{\rm F}(x,t)=-\mathrm{sgn}(e) \hbar v_{0}\sqrt{\pi n(x,t)}$~\cite{Rudin2011}, $E(x,t)$ is the in-plane electric near field.
Equations  (\ref{A4_1}) and (\ref{A4_2}) are taken from Ref. \cite{Rudin2011} by approximating the carrier momentum $p(x,t)$ for small carrier velocities, $v(x,t)<v_{\rm F}$, by 
\begin{equation*}
p(x,t)=-\mathrm{sgn}(e) \frac{\epsilon_{\rm F}(x,t)}{v_{0}^2}v(x,t).
\end{equation*}
Strictly speaking, the latter equation is valid for zero temperature. However, as mentioned in Section~\ref{theory_1}, the free-carrier gas in doped (or gated) graphene is degenerate even at room temperature so that this expression is relevant also for room temperature. We also neglect the terms describing the carrier pressure and viscosity contributions in Eq. (\ref{A4_2}) which are responsible for the non-locality effects in the plasmonic response.

 Nonlinearity of the free carrier motion in graphene described by Eqs.~(\ref{A4_1}) and~(\ref{A4_2}) originates from (i) the product $n(x,t)v(x,t)$ defining the conductive current  $j(x,t)$ ,  (ii) the second term in Eq. (\ref{A4_2}) describing the nonlinear convection current, and (iii) the dependence of the oscillating Fermi energy $\epsilon_{\rm F}(x,t)$ on the applied electric-field amplitude $E(x,t)$. It is worth noting that all the three sources of the nonlinearity survive only if an inhomogeneous (i.e., the $x$-dependent) carrier-density oscillations occur in graphene. Therefore, these nonlinearities essentially are of the plasmonic nature. 

We solved the hydrodynamic equations (\ref{A4_1}) and (\ref{A4_2}) in the perturbation approach \cite{Aizin2007} by expanding every unknown variable in powers of the electric field amplitude and keeping only linear and quadratic terms in this expansion. Then the induced current density in graphene is given by  $j(x,t)=en_0 v_1(x,t)+e n_1(x,t) v_1(x,t)$ , where $n_0$ is the equilibrium carrier density, and $n_1(x,t)$ and $v_1(x,t)$ are the linear corrections to the density and velocity of free carriers in graphene, respectively.

Time averaging of $j(x,t)$ yields the rectified current
\begin{equation}
j_{\rm{pl}}=-\frac{2 e^3 \tau^2 v_{0}^2}{\pi \hbar^2 \omega (\tau^2 \omega^2+1)} \sum_q q(|E_q|^2-|E_{-q}|^2),   
\label{A4_4}
\end{equation}
where $E_q$ are the amplitudes of the spatial Fourier harmonics of the plasmonic electric field $E(x)$, $q=2\pi l / d$ and $l$ is an integer. It follows from Eq. (\ref{A4_4}) that the {\it dc} photocurrent appears only for $E_q\ne E_{-q}$, due to the differential drag of the carriers by the counter-directed Fourier harmonics of the plasmonic near field. The differential plasmonic photocurrent has the opposite polarities depending on the electron or hole conductivity of graphene. 
In distinction from conventional 2D electron system~\cite{Popov2015}, the prefactor in the sum~(\ref{A4_4}) is independent of the equilibrium  carrier density which means that Eq.~(\ref{A4_4}) for the differential plasmonic drag current  is valid for both a homogeneous and periodically modulated graphene. However, additional contributions to the plasmonic ratchet photocurrent, which can appear in graphene with spatially modulated carrier density, requires further analysis.

\section*{Acknowledgments}
Financial  support via the Priority Program 1459 Graphene of the German Science Foundation DFG 
and the Russian Foundation for Basic Research are gratefully acknowledged.


\begin{thebibliography}{99}
\bibitem{footnote3}  While being detected in all reported measurements a 
polarization independent offset given by the coefficient $V_{\rm off}$ 
will not be discussed in details. Instead, hereafter we focus on 
helicity  sensitive photocurrent, $j_{\rm C} \propto V_{\rm C}$,
and currents driven by linearly polarized light, 
$j_{L1} \propto V_{\rm L1}$ and $j_{L2} \propto V_{\rm L2}$.
\bibitem{footnote2} Note that 
to measure the circular effect we would need to detect current in the $y$-direction. However, necessity
of the gate contact does not allow to fabricate them.
\bibitem{footnote4}
Note that due to the addressed above hysteresis in the back gate sweep the exact 
assignment of the photocurrent nodes and maxima with respect to the Dirac point is not possible.
\bibitem{footnoteLG} Here we assume the Boltzmann statistics for clarity. The case of the Fermi-Dirac statistics and real situation of degenerate carriers are studied in Sec.~\ref{theory_all} and Appendix~\ref{A2}.
\bibitem{footnoteT1}
Hereafter we consider a graphene sheet with the lateral potential ${\cal V}(x)$. 
The electron energy in each valley, $K$ or $K'$, is given by
%
$\varepsilon_{\bm k} = \hbar v_0 k + {\cal V}(x)$,
%
and the two-dimensional wave vector ${\bm k}$ is referred to the vortex of the hexagonal Brillouin zone. Since in the model under consideration the behavior of electrons in the $K$ or $K'$ valleys is identical we consider the current generation in one of them and then double the result.
\bibitem{footnoteStokes} Note that the response to the polarized radiation may change
due to the presence of the grating-gate polarizer as discussed in details below, see Sec.~\ref{theory_all}B.
\bibitem{footnoteSeebeck} Note that while 
the Seebeck thermoratchet photocurrent itself 
does not depend on the radiation polarization. Its magnitude
in response to polarized radiation may change
due to the presence of the grating-gate 
polarization effect.
\end{thebibliography}
\end{document}